\documentclass[tightenlines,preprint,amsmath,amssymb,aps]{revtex4-1}
\usepackage[utf8]{inputenc}
\usepackage[english]{babel}
\usepackage{xcolor}
\usepackage{braket}
\usepackage{bm,amsmath}
\usepackage{amssymb}
\usepackage{graphicx}
\usepackage{mathtools}
\usepackage[normalem]{ulem}
\newcommand{\si}{\sigma}
\newcommand{\da}{^{\dagger}}
\newcommand{\ep}{\varepsilon}
\newcommand{\om}{\omega}
\newcommand{\Om}{\Omega}
\newcommand{\la}{\lambda}
\newcommand{\upaw}{\uparrow}
\newcommand{\dnaw}{\downarrow}

\newcommand{\B}{\textcolor{blue}}

\begin{document}
     \title{Effective Hamiltonians in the Quantum Rabi Problem}
\author{P. Gartner$^1$}
\author{V. Moldoveanu$^{1}$}
%\email[Corresponding author: ]{valim@infim.ro}
\affiliation{$^1$ National Institute of Materials Physics, PO Box MG-7, Bucharest-Magurele, Romania}
    
    \date{\today}

\begin{abstract}

We revisit the theoretical description of the ultrastrong light-matter interaction in terms of exactly solvable 
effective Hamiltonians. A perturbative approach based on polaronic and spin-dependent squeezing transformations
provides an effective Hamiltonian for the quantum Rabi model up to the second order in the expansion parameter. 
The model consistently includes both rotating and counter-rotating terms, going therefore beyond the 
rotating wave approximation. Analytical and numerical results show that the proposed Hamiltonian performs better than 
the Bloch-Siegert model when calculating operator averages (e.g.\, the mean photon number and number of excitations). 
This improvement is due to a refined calculation of the dressed states within the present model. 
Regarding the frequency shift induced by the qubit-photon interaction, we find a different sign from the Bloch-Siegert value. 
This influences the eigenstates structure in a non-trivial way and ensures the correct calculation of the number of
excitations associated to a given dressed state. As a consistency check, we show that the exactly solvable independent boson model 
is reproduced as a special limit case of the perturbative Hamiltonian.

\end{abstract}  
%\pacs{42.50.Pq, 03.65.Yz, 42.50.Lc}      
  \maketitle
  
\section{Introduction}

The physics of strongly coupled qubit-photon hybrid systems has become a steadily developing field \cite{Kockum,RMP} which
 challenges the existing theoretical methods. In particular, as the field-matter coupling strength 
increases, the Jaynes-Cummings model \cite{JCM}, which neglects the so-called off-resonant transitions 
within the rotating wave approximation (RWA), become\B{s} less accurate.

More general models taking fully into acount the interaction between a two-level system (e.g.\, a superconducting qubit 
or a quantum dot) and a boson mode (cavity photon or transmission line resonator) are defined by the class of Hamiltonians
\begin{equation}
  H = \frac{1}{2}\ep\si_z + \om a\da a +g_1(a\da\si+a\si\da) + g_2 
(a\da\si\da+a\si) \, ,
  \label{eq:hg1g2}
\end{equation}
 where the quasi-spin notation is used for the qubit and $a,a\da$ are bosonic operators of the single-mode field. 
The corresponding energies are $\ep$ and $\om$ respectively, and $g_1$ and $g_2$ are the coupling strengths for the resonant (rotating)
 and off-resonant (counter-rotating) processes. We take $\hbar=1$ throughout. The Jaynes-Cummings (JC) Hamiltonian is then recovered 
from Eq.(\ref{eq:hg1g2})
by taking $g_2=0$, while the quantum Rabi model (QRM) \cite{Meystre,Walls-Milburn} corresponds to $g_1=g_2=g$:
\begin{equation}
   H_R = \frac{1}{2}\ep\si_z + \om a\da a +g \si_x (a\da+a)  \, .
  \label{eq:hqrm}
\end{equation}

Generally, the problems defined by Eq.\eqref{eq:hg1g2} are not exactly soluble, one notable exception being
the quantum Rabi Hamiltonian whose spectrum was calculated \cite{Braak1,Braak2,Wang} as zeroes of recursively defined new
special functions or in terms of confluent Heun functions \cite{Zhong}. The mathematics is quite involved both for the 
eigenvalues and for eigenvectors.
In view of this situation, in the literature one prefers to devise
effective Hamiltonians \cite{RMP} to obtain approximate eigenvalues and eigenvectors, in suitably restricted domains of the parameter
space defined by $\{\ep,\om, g \}$.
Also, simple, analytic effective descriptions are strongly preferred when including additional interactions to
describe optical, transport \cite{Cirio,DMG} or relaxation processes \cite{DeLiberato,Beaudoin,Settineri}.

A useful test for these approximations consists in inspecting limit cases in which exact solutions are known for the full
 Hamiltonian. One such case is the JC model mentioned above, in which the RWA brings the Hamiltonian to a $2 \times 2$ 
block-diagonal form. A second one is the independent boson model (IBM) \cite{Mahan} which is extensively used for quantum optical 
and transport calculations involving carrier interaction with photons, phonons \cite{Krummheuer} or vibrons \cite{Mitra}. 
In the present case the IBM limit is obtained by taking $\ep=0$ in the QRM, Eq.~\eqref{eq:hqrm}, and the corresponding Hamiltonian is 
diagonalized by a displacement of the bosonic oscillator centers brought in by the Lang-Firsov, or polaronic unitary transform. 

In deriving approximate effective Hamiltonians for the quantum Rabi problem one usually employs a limited expansion 
of various unitary transforms in terms of a suitable small parameter. An example is provided by the dispersive Hamiltonian 
\cite{Zueco, Blais} obtained by a unitary transform aiming at diagonalizing the qubit-boson interaction, expanded to second order in 
the parameter $\la = g/|\ep-\om|$, in the weak coupling limit $\la \ll 1$. 
The resulting effective Hamiltonian reproduces the JC spectrum to second order in $\la$, i.e. within 
the domain of parameters in which it was derived. Moreover, it recovers the exact result in the IBM 
limit.  

Another popular effective Hamiltonian is the Bloch-Siegert (BS) one \cite{Klimov-Chumakov,Rossatto}. It 
treats the QRM problem up to second order in the parameter $\la=g/(\ep+\om)$ and brings it to a renormalized 
form of the JC Hamiltonian. Two succesive unitary transforms are used in the process and higher order off-resonant terms are discarded 
as rapidly oscillating, in the spirit of the RWA. 
Supposedly, the validity domain of this approach is roughly $g/\om \lesssim 0.3$ and corresponds to the so-called perturbative 
ultra strong coupling regime (pUSC) \cite{RMP}. The perturbative criteria for the BS Hamiltonian was recently used 
\cite{Rossatto} to establish a spectral classification of the field-matter coupling regimes in the QRM. On the experimental side the 
Bloch-Siegert shift was recently observed in strongly coupled qubit-oscillator systems \cite{Diaz-BS}. An alternative improvement of
the JC model relies on the Van Vleck perturbation theory to the second order in the qubit-field coupling \cite{Hausinger}.

Surprisingly, a simple analysis shows that the IBM limit is {\it not} recovered by the BS Hamiltonian, despite the fact 
that taking $\ep=0$ does not preclude the expansion parameter, now equal to  $g/\om$, to remain small. Still, the BS model 
is tacitly accepted as the main candidate to describe the ultra-strong coupling regime. 

In view of this situation, in what follows we show how to obtain an effective Hamiltonian for 
the QRM, in the same perturbative conditions as for the BS model, i.e. up to the second order in $\la$, having 
the same simple, renormalized JC structure and which additionally meets the test 
of recovering the exact IBM result for $\ep=0$. 
Moreover, it will become obvious that the procedure is {\it purely perturbative}, i.e. it does not have to 
 rely on discarding off-resonant terms. 

As the qubit energy $\ep$ approaches resonance the spectra of the two effective models become more similar. 
Nevertheless, we show that important differences remain in the dressed states and  
are inherited by the expectation values provided by these states. Especially the excitation number turns out to be sensitive to 
these differences.

The rest of the paper is organized as follows. The perturbative effective Hamiltonian is derived in Section II. 
A detailed comparison of this perturbative Hamiltonian with the Bloch-Siegert model is presented and discussed in Section III. 
Their spectra, average photon number, population inversion and number of excitations are obtained and compared both analytically and 
numerically in subsections III A, III B, and III C. We also compare our result with the so called generalized 
rotating wave Hamiltonian (GRWA) obtained by Irish \cite{Irish} in the deep strong coupling case (subsection III D). 
This regime is defined by large values 
of the coupling, $g/\om \approx 1$ and beyond, but the numerical evidence \cite{Yu,Xie,Zhang2016} shows that this approach 
gives good results at lower couplings too. Therefore comparison against it is relevant. Besides, the GRWA Hamiltonian meets 
the IBM limit test. Section IV is left for conclusions.

\section{An IBM-compatible effective Hamiltonian}

We consider unitary transforms of the form $e^S$, with $S\da=-S$ and their expansion in iterated 
commutators with their generators $S$. Then any operator $A$ transforms as
\begin{equation}
A_S=  e^S A\, e^{-S}  = A + \left[S,A\right] + 
                         \frac{1}{2!}\left[S,\left[S,A\right]\right] + \dots
\label{eq-Utf}                         
\end{equation}

Now let us rewrite the Rabi Hamiltonian of Eq.\eqref{eq:hqrm} as
\begin{equation}
H_R = \frac{1}{2}\ep\si_z + H_{IBM}\, ,
  \label{eq:hqrm2}
\end{equation}
in which we separate the free qubit contribution and the independent boson model part.
The linear part in $H_{IBM}$ can be absorbed in the quadratic term by a shift of the bosonic operators, produced by the generator
\begin{equation}
  S = \la (a\da -a)\, \si_x \,
  \label{eq:lf}
\end{equation}
with the value of $\la$ to be chosen later.
Using
\begin{equation}
 [S,a\da] = - \la\, \si_x \, , \qquad [S,a] = - \la\,\si_x \, ,
\end{equation}
it is clear that the canonical transformation amounts to a simple shift  
\begin{equation}
 a\da_S = a\da - \la\,\si_x \, , \qquad a_S = a - \la\,\si_x \, .
 \label{eq:a_shift}
\end{equation}
As a result, the transformed Hamiltonian becomes
\begin{equation}
H_{IBM,S} = \om a\da a + (g-\la \om)(a\da+a)\si_x +(\la^2\om -2\la g)\, .
\end{equation}
Choosing $\la = g/\om$ removes the interaction term and diagonalizes the IBM 
Hamiltonian. 
\begin{equation}
 H_{IBM,S}= \om a\da a -g^2/\om\, .
\label{eq:H-IBM}
\end{equation}
Therefore the IBM problem alone is exactly solvable. The solution 
consists of the oscillator keeping its frequency $\om$, but shifted by $g/\om$,  
the shift direction being decided by the eigenstate of $\si_x$

The presence of the free qubit part in Eq.\eqref{eq:hqrm2} spoils this simple scenario. 
Both the frequency and the shift are modified, and we show below that in these 
circumstances a different choice for the value of $\la$ is more appropriate.

Obviously, $\si_x$ is left unchanged by the IBM transform. In contrast, the 
change in $\si_z$ is rather complex. One can easily check that:
\begin{align}
[S, \si_z] = & - 2 \,\la (a\da-a)\, i\si_y \,, \nonumber \\
[S,i\si_y] = & - 2 \,\la (a\da-a)\, \si_z \, .
\label{eq:Ssi_z}
\end{align}
This means that successive commutators generate alternatively the $\si_z$ and 
$i\si_y$ operators, and each step brings in an additional factor $-2\, \la 
(a\da-a)$. The $\si_z$ operator collects the even 
terms, which add up to a hyperbolic cosine function, while the $i\si_y$ terms 
generate a hyperbolic sine function. 
The net result is
\begin{align}
& H_{R,S} =  \om a\da a +  (g-\la \om)(a\da+a)\si_x +(\la^2\om -2\la g) \nonumber 
\\
& +\, \frac{1}{2} \ep \left \{\cosh[2\, \la(a\da-a)]\, \si_z 
                        -\sinh[2\, \la(a\da-a)]\,i\si_y \right \} \,. 
  \label{eq:no_expansion}
\end{align}
Up to now no approximation is involved. A simpler Hamiltonian, valid to 
the second order in $\la$, is obtained by expanding 
$\cosh(x) \approx 1+x^2/2$ and $\sinh(x) \approx x$. 
Taking into account that $\si_x=\si\da + \si$ and $i\si_y=\si\da - \si$, 
this expansion leads to an effective Hamiltonian $H'$ of the form 
\begin{align}
& H_{R,S}\approx  H' = \frac{1}{2}\ep \si_z + \om a\da a + (\la^2\om -2\la g) \nonumber \\
        & + (g-\la\om+\la\ep) (a\da \si +a \si\da) 
          + (g-\la\om-\la\ep) (a\da \si\da +a \si) \nonumber \\
        & + \la^2 \ep (a\da -a)^2 \si_z \, .
\end{align}
Now it is clear that an appropriate choice for $\la$ is
\begin{equation}
\la = \frac{g}{\ep+\om} \,
\label{eq:lambda}
\end{equation}
which cancels out the off-resonant term linear in the bosonic operators. This 
is the same small parameter as the one used in the Bloch-Siegert formalism 
\cite{Klimov-Chumakov}.
The effective Hamiltonian has now the expression
\begin{align}
H' & = \frac{1}{2}\ep \si_z + \om a\da a 
            - (2\, \la g -\la^2 \om) \nonumber \\
& + 2\la\ep (a\da \si +a \si\da) + \la^2 \ep (a\da - a)^2 \si_z \, .
\label{eq:heff}
\end{align}

Finally, one has to handle the last, quadratic term. To this end one 
considers a second unitary transform, with the generator
\begin{equation}
S' = \frac{1}{2} \eta (a^{\dagger \,2}-a^2)\si_z \, .
\end{equation}
This $\si_z$-dependent squeezing transform \cite{Comm-Zhang} leads to
\begin{equation}
 [S',a\da] = - \eta \,a \, \si_z \, , \qquad [S',a] = - \eta \,a\da \si_z \, ,
\end{equation}
in which again succesive commutators generate alternatively the same two 
operators. One gets a Bogolyubov-type transform
\begin{align}
 a\da_{S'} & =\cosh \eta\, a\da -\sinh \eta \, a\, \si_z \, ,\nonumber \\
 a_{S'}&= -\sinh \eta \, a\da\, \si_z + \cosh \eta \, a \, .
\end{align}
Different Hamiltonian pieces change accordingly, for instance the photon number operator 
\begin{align}
& a_{S'}\da a_{S'} =  \cosh 2\eta\, a\da a + \sinh^2 \eta \nonumber \\
           & - \frac{1}{2} \sinh 2\eta (a\da-a)^2 \si_z 
             - \frac{1}{2} \sinh 2\eta (2 a\da a +1)\si_z\, , 
             \label{eq:quadr1}
\end{align}
and
\begin{align}
  (a_{S'}\da - a_{S'})&= (\cosh \eta+\sinh\eta \,\si_z)(a\da -a) \,, \nonumber \\
   (a_{S'}\da - a_{S'})^2\si_z &= [\cosh 2\eta \, \si_z +\sinh 2\eta] (a\da -a)^2 \, .
  \label{eq:quadr2}
\end{align}
From Eqs.\eqref{eq:quadr1},\eqref{eq:quadr2} above the quantities in $H'_{S'}$ containing $(a\da -a)^2$ are identified as:
\begin{eqnarray}
X_z&:=& \si_z\left (\la^2 \ep \cosh 2\eta -\frac{\om}{2} \sinh 2 \eta\right ) (a\da -a)^2 \, ,\\\label{eq:eta}
X_0&:=&  \la^2 \ep \sinh 2\eta(a\da -a)^2 \, .
\end{eqnarray}
The terms contained in $X_z$ cancel out by choosing $\eta$ so that
\begin{equation}
\frac{1}{2}\sinh 2 \eta = \frac{\la^2 \ep}{\om} \cosh 2\eta \, . 
\end{equation}
The condition makes $\eta$ of second order in $\la$ and, neglecting higher 
contributions, one has 
\begin{equation}
\eta = \frac{\la^2 \ep}{\om} \, .
\label{eq:eta2}
\end{equation}
In this case the term $X_0$ in Eq.\eqref{eq:eta} is of fourth order in $\la$ and is discarded. 

The terms in the Hamiltonian Eq.\eqref{eq:heff} not considered up to now are 
the free qubit energy, which commutes with $S'$, and the JC interaction 
$2\la\,\ep (a\da \si +a \si\da)$, which is already a first order quantity, and 
therefore its transformation by $S' \sim \la^2$ brings in changes of at least 
third order. Therefore both are kept unchanged.

Finally, expanding systematically the hyperbolic functions to second order in 
$\la$ (i.e. first order in $\eta$) 
and spelling out the expressions of $\la$ and $\eta$ in terms of the 
model parameters, leaves us with $ H'_{S'}\approx H_P $ where
the perturbative IBM-compatible effective Hamiltonian is given by
\begin{align}
 H_P= &\frac{\ep}{2} \left(1 -\frac{2g^2}{(\ep+\om)^2} \right)\si_z 
            +\left (\om-\frac{2g^2\ep}{(\ep+\om)^2}\,\si_z \right)\, a\da a 
                                                       \nonumber \\
           & +  2\,\frac{g\,\ep}{\ep+\om} (a\da \si+a\, \si\da) 
             - \frac{g^2}{\ep+\om}\, \frac{2\,\ep +\om}{\ep+\om}\, .
             \label{eq:heff2}
\end{align}
This expression of the effective Hamiltonian $H_P$ is the main result 
of the paper.

It is now obvious that for $\ep=0$, $H_P$ recovers exactly the 
IBM result, Eq.\,(\ref{eq:H-IBM}) and, moreover, this holds for all values of the coupling constant $g$. 

\section{Comparing effective Hamiltonians}

The Bloch-Siegert and the perturbative effective Hamiltonian 
given by Eq.\,(\ref{eq:heff2}) have both the  structure of a renormalized JC model
\begin{equation}
H = \frac{1}{2} (\ep + \Om)\si_z + (\om +\Om \si_z)a\da a 
                           + \gamma(a\da \si+a \si\da)-\delta \, .
   \label{eq:heff_common}
\end{equation}
The difference comes from the parameters $\Om, \gamma$ and $\delta$. They are given 
by \cite{Rossatto}
\begin{equation}
\Om = \frac{g^2}{\om +\ep}, \quad \gamma = g, \quad \delta=\frac{1}{2}\, 
                                    \frac{g^2}{\om+\ep} \,
    \label{eq:param_BS}
\end{equation}
for $H=H_{BS}$, and by
\begin{equation}
\Om = -\frac{g^2}{\om +\ep}\,\frac{2\ep}{\om+\ep}, \quad \gamma = g\,\frac{2\ep}{\om+\ep}, 
  \quad \delta= \frac{g^2}{\om+\ep} \,\frac{\om +2\ep}{\om+\ep} \,
  \label{eq:param_P}
\end{equation}
    
for $H=H_P$. 

The JC structure of Eq.\eqref{eq:heff_common} reduced the problem to $2\times 2$ blocks, 
indexed by the photon number $n$. In the corresponding basis 
$\left\{ \ket{n,\upaw}, \ket{n+1,\dnaw} \right \}$ the $n$-th block has the form
\begin{eqnarray}\nonumber
{\hat h}_{n}&=&\alpha_n \si_0+\frac{1}{2}\Delta_n \si_z+\gamma\sqrt{n+1} \si_x\\\label{eq:al_tht}
&=&\alpha_n \si_0 + R_n (\cos \theta_n \si_z + \sin \theta_n \si_x)\, ,
\end{eqnarray}
where $\si_0$ is the $2\times 2$ unit matrix and we introduced
\begin{eqnarray}
\alpha_n&=&\om\, (n+1/2) -\left(\frac{\Om}{2} +\delta\right) \\
R_n &=& \sqrt{\Delta_n^2/4 +\gamma^2(n+1)} \, ,
\end{eqnarray}
with $\Delta_n = \ep -\om +2\,\Om\, (n+1)$ acting as an effective, $n$-dependent detuning. The angle
$\theta_n$ is defined by the relations $\cos \theta_n= \Delta_n/(2R_n)$ and 
$\sin \theta_n = \gamma \sqrt{n+1}/R_n$.  Note that $\Om/2 + \delta$ turns out to take the 
same value $g^2/(\om +\ep)$ in both effective Hamiltonians considered, therefore
the middle points of their spectra coincide.

It is known that the spectrum of ${\hat h}_{n}$ is $E_{n,\pm}=\alpha_n \pm R_n$ with the 
eigenvectors (i.e.\,dressed states) given by the linear combination
\begin{equation}
\ket {\varphi_{n,\pm}}:=u_{\pm} \ket {n, \upaw} + v_{\pm} \ket{n+1,\dnaw} 
\label{eq:eigvec}
\end{equation}
where $(u_+,v_+) =(\cos \theta_n/2, \sin \theta_n/2)$ for the higher state $\ket {\varphi_{n,+}}$ and 
$(u_-,v_-) =(\sin \theta_n/2, - \cos \theta_n/2)$ for the lower state $\ket {\varphi_{n,-}}$. To complete 
the picture we also {calculated give the energy of the ground state $\ket{0, \dnaw}$ 
as $E_0=-\ep/2 - (\Om/2+\delta)$.

By examining Eqs.\eqref{eq:param_BS} and \eqref{eq:param_P}, it is clear that the differences
between the two models are more pronounced for small values of $\ep$. As already mentioned, at 
$\ep=0$, $H_P$ recovers the exact IBM result, while $H_{BS}$ does not. 

On the other hand at resonance, $\ep=\om$, the parameters of the two models coincide, {\it except for 
the sign of} $\Om$. The latter has no bearing on the spectra, which therefore are the same for 
the two models, but it modifies the structure of the eigenstates. Indeed, at resonance the sign 
of $\Om$ carries over to a sign change in $\cos\theta_n$ which in turn leads to exchanging $\cos \theta_n/2$
and $\sin\theta_n/2$. 
In other words, even though at resonance the spectra of $H_{BS}$ and $H_P$ coincide, because of the
sign difference in $\Delta_n$ the attribution of the eigenvalues is flipped:
for example, the weight of the two basis vectors $\ket {n, \upaw}$ and $\ket{n+1,\dnaw}$ in the 
upper state $\ket {\varphi_{n,+}}$ of the BS Hamiltonian is the same as for the lower state 
$\ket {\varphi_{n,-}}$ of the perturbative Hamiltonian. Further consequences of this fact are 
analyzed in subsection \ref{subsC}.

To resume the situation, the two models predict different spectra and states, as suggested by these
two examples. Intermediate values for $\ep \in [0,\om]$ are expected to lead to similar conclusions.
Let us also remark that $\Om$ is interpreted as a 
bosonic frequency shift which depends on the qubit state, and detecting the former gives 
information about the latter \cite{RMP}. Therefore the sign difference means that this way
of reading the qubit state works in opposite directions in the predictions of the two models.

Still, these arguments are merely heuristic. Eigenstates are not directly comparable, $H_P$ and 
$H_{BS}$ having been obtained by different unitary transforms, i.e. the two effective Hamiltonians 
work in different bases. The relevant comparison is on basis-independent quantities, to which we turn below.

\subsection{Energy spectra}

\begin{figure}[t]
        \includegraphics[angle=0,width=0.45\textwidth]{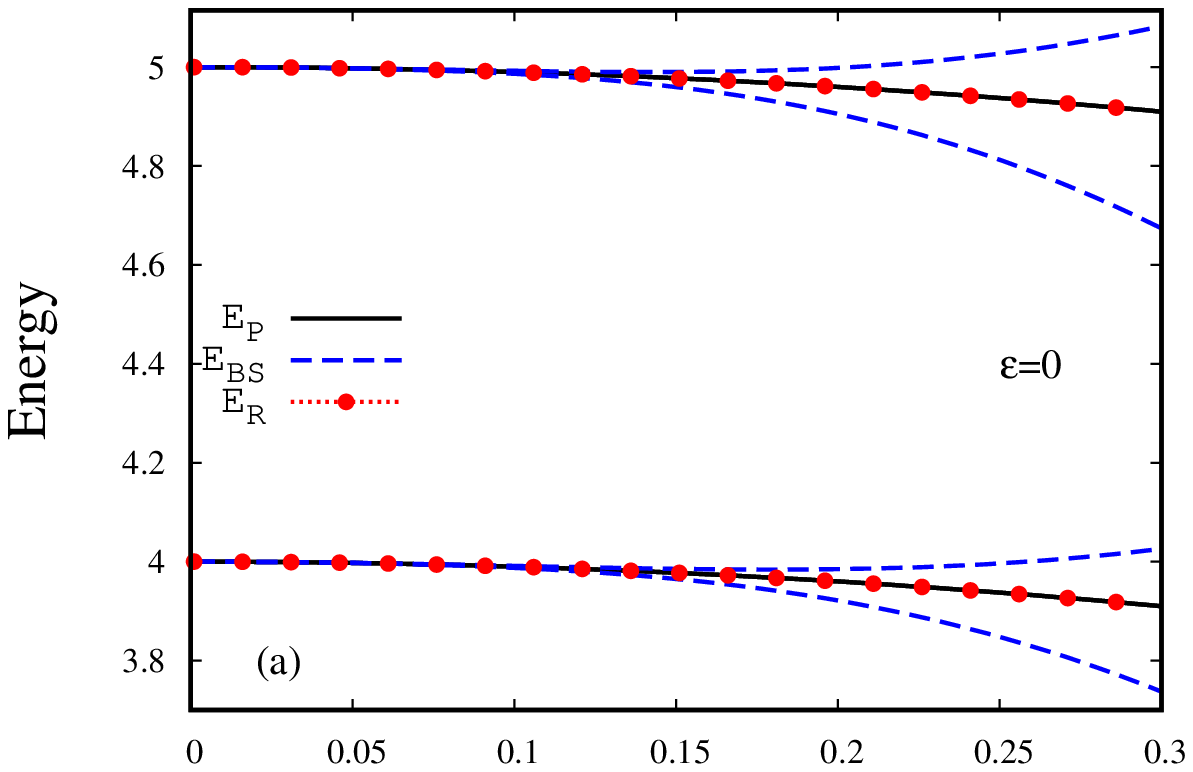}
	\includegraphics[angle=0,width=0.45\textwidth]{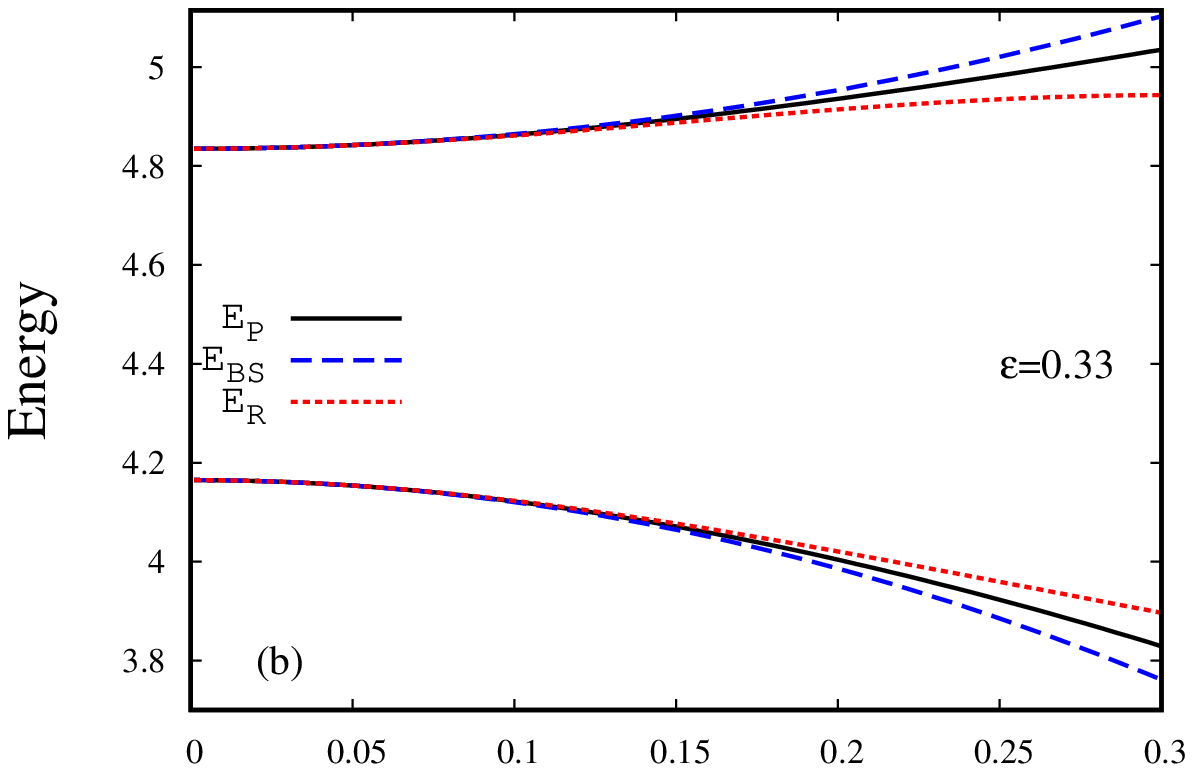}
	\includegraphics[angle=0,width=0.45\textwidth]{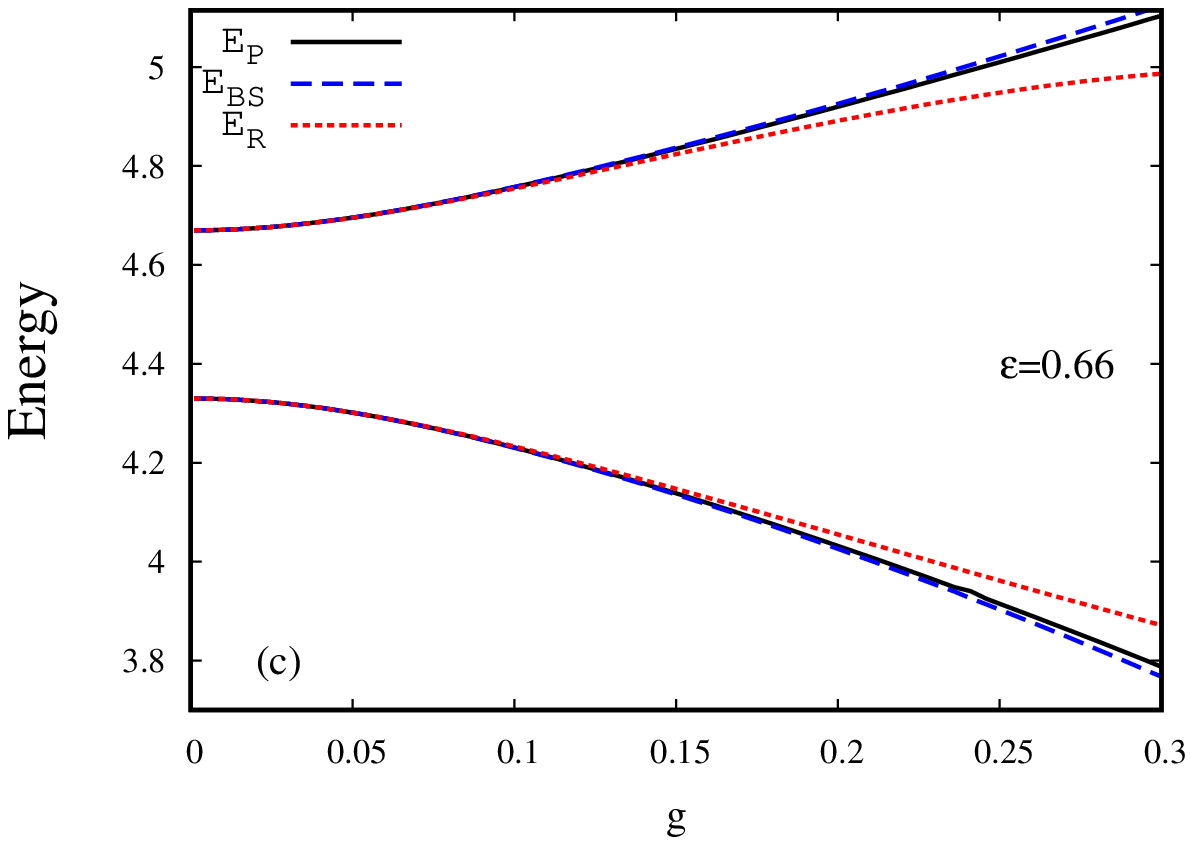}
        \caption{Two spectral branches of the Rabi (R), Bloch-Siegert (BS) and perturbative (P) Hamiltonians as a function of
        the matter-photon coupling $g$ at different values of the qubit energy:
         (a) $\ep=0$, (b) $\ep=0.33$ and (c) $\ep=0.66$. The spectrum of the QRM is calculated numerically.
         $g$ is restricted to the perturbative USC regime. Panel (a) also includes the branches $E_{3,+}$ and $E_{5,-}$ 
	 (see the text). }
\label{Spec1}
  \end{figure}
  At resonance, $\ep=\om$, all parameters in Eq.\eqref{eq:al_tht} are exactly the same for the two cases, except
for sign of $\Delta_n$, and thus the spectra are identical. On the other hand, in the limit $\ep=0$, $H_P$ becomes exact.
Therefore, it is expected that for the whole subresonant range, $\ep \leqslant \om$, our approach is an improvement over
the BS efective Hamiltonian.

To illustrate this fact we compare in Fig.\,\ref{Spec1} the spectra of the two effective models (BS and perturbative),
and those produced by numerical diagonalization of the full Rabi Hamiltonian. For the latter a cutoff at $N_{{\rm ph}}=30$
photons is found to be sufficient. We use $\om =1$ units throughout, since the problem is homogeneous of degree $1$ in
the parameters, i.e. it is defined essentially by only two dimensionless parameters $\ep/\om$ and $g/\om$.

We selected the pair of spectral branches $E_{4,\pm}$ for three values of the qubit energy $\ep$. 
However, in the IBM limit (i.e\, for $\varepsilon=0$, see Fig.\,\ref{Spec1}(a)) two additional branches are added, to illustrate
 that the coinciding spectra of the Rabi and perturbative Hamiltonians are doubly degenerate, while the Bloch-Siegert spectrum 
 displays a degeneracy lifting as the qubit-photon coupling strenght increases. Note that for the sake of clarity the spectral 
 branches of $H_P$ are represented by red circles. 

 To understand these differences one has to recall that in the limit $\ep=0$, $H_P$ becomes exact, while the BS spectra are correct 
 only to $\la^2$ accuracy, as it is easy to check on the expression for $E_{n, \pm}$. 
With increasing $\ep$ the spectral difference between the models diminishes, see Fig.\, \ref{Spec1}(b),(c), as predicted by the
analytic expression of $H_P$ and $H_{BS}$ until, for $\ep =1$ they coincide (not shown). In the process the perturbative values
remain closer to the exact Rabi result than the BS ones.

\subsection{Operator averages. Photon population}

When calculating the expectation value of an operator on a given eigenstate of the 
effective Hamiltonian, one has to take into acccount the unitary transforms that 
lead to it. The operators should be 'rotated' by the same procedure as the 
Hamiltonian.

In the case of $H_P$, we use again the two generators $S=\la(a\da-a)\si_x$ and 
$S'= \frac{1}{2}\eta(a^{\dagger\, 2}-a^2)\si_z$, with the parameters given in Eqs.\eqref{eq:lambda} and \eqref{eq:eta2}.
The observable we are interested in is the photon number $\hat{n}=a\da a$. From Eq.\eqref{eq:a_shift}
we obtain by the first transform
\begin{equation}
a \da_S a_S = a\da a - \la (a \da +a)\si_x +\la^2  \, .
\label{eq:Sn}
\end{equation}
The second step involves $S'$ and is apparently more complicated but it gets simpler by limiting 
the expansion to second order in $\la$ and taking into account that $\eta \sim \la^2$.
For instance, it is clear that the second term in Eq.\eqref{eq:Sn} remains unchanged. 
Indeed, changes are at least of first power of $\eta$ and with the prefactor $\la$ they 
are third order quantities. 

For the first term $a\da a$ one uses Eq.\eqref{eq:quadr1} expanded to second order in $\la$, with the net result that 
after both transforms one has
\begin{equation}
\hat{n}_{S,S'}\approx\hat{n} - \eta(a^{\dagger\, 2}+a^2)\si_z - \la (a \da +a)\si_x +\la^2  \, .
\label{eq:SprimSn}
\end{equation}
This is the operator to be averaged on the eigenstates of the form given in 
Eq.\eqref{eq:eigvec}. Things turn out to be further simplified by noting that 
operators like $a^2, a \si$ and their hermitic conjugates do not contribute to 
the expectation value, since they do not connect vectors belonging to the same two-dimensional subspace.

To conclude, one has
\begin{align}
\left<\hat{n}\right>_{P,\pm} = &\bra {\varphi_{n,\pm}} 
                     [a\da a -\la (a\da \si + a \si\da) +\la^2]
                     \ket {\varphi_{n,\pm}} \nonumber \\   
                     = & u^2_{\pm} n + v^2_{\pm}(n+1) -2 \la u_{\pm} v_{\pm} \sqrt{n+1} +\la^2  \, ,
\label{eq:av_P}
\end{align}
where the index $P$ recalls that the average is calculated w.r.t. the dressed states of $H_P$. Obviously, 
$\left<\hat{n}\right>_{P,\pm}$ depends on the specific two-dimensional subspace $n$ as well but we drop 
this additional index for the simplicity of writing.
Replacing $u_{\pm}$ and $v_{\pm}$ leads to the values
\begin{align}
\left<\hat{n}\right>_{P,\pm}=\, n + \frac{1}{2}(1\mp\cos\theta_n) \mp \la \sin\theta_n \sqrt{n+1} + \la^2 \, .
\label{eq:aver_P}
\end{align}
A negative sign of $\Om$ means a lower $\cos\theta_n$ value, and this encourages a higher 
photon population in the upper state. 
For the ground state one obtains $\left<{\hat n}\right>_{P,0} = \la^2$.

The prediction of the Bloch-Siegert Hamiltonian for the same quantity is evaluated by the same procedure.
Some differences arise though, beginning with the first generator which now reads
$\tilde{S}=\la(a\da \si \da- a \si)$ with the result that now
\begin{equation}
a_{\tilde{S}}\da a_{\tilde{S}} = a\da a - \la (a \da \si \da +a \si) +\la^2 \si \si \da 
                        - \la^2 a\da a\si_z\, .
\label{eq:Sn_BS}
\end{equation}
The second transform has the same form $\tilde S'= \frac{1}{2}\eta(a^{\dagger\, 2}-a^2)\si_z$
with $\eta\sim\la^2$. 

As before, the terms containing $\la$ remain unchanged 
up to the second order, and the first term $a\da a$ is modified by a combination of
$a^2$ and $a^{\dagger\,2}$, which do not contribute to the expectation value. The new
feature is that no contribution comes from the second term in Eq.\eqref{eq:Sn_BS} above.  
Therefore one is lead to
\begin{align}
\left<\hat{n}\right>_{BS,\pm} = &\bra {\varphi_{n,\pm}}
                    [a\da a +\la^2 \si \si \da -\la^2 a\da a \si_z]
                      \ket {\varphi_{n,\pm}} \nonumber \\
                      = & \, u^2_{\pm} (n-\la^2 n) + v^2_{\pm}(n+1+\la^2(n+2))\, .                        
\label{eq:av_BS}
\end{align}
As a result one has
\begin{align}
\left<\hat{n}\right>_{BS,\pm}=\, n + \frac{1}{2} +\la^2 
                \mp \left(\frac{1}{2}+\la^2(n+1)\right) \cos\theta_n \, .
\label{eq:aver_BS}
\end{align}
For the ground state one finds $\left<\hat{n}\right>_{BS,0} = \la^2$.

\begin{figure}[t]
        \includegraphics[angle=0,width=0.45\textwidth]{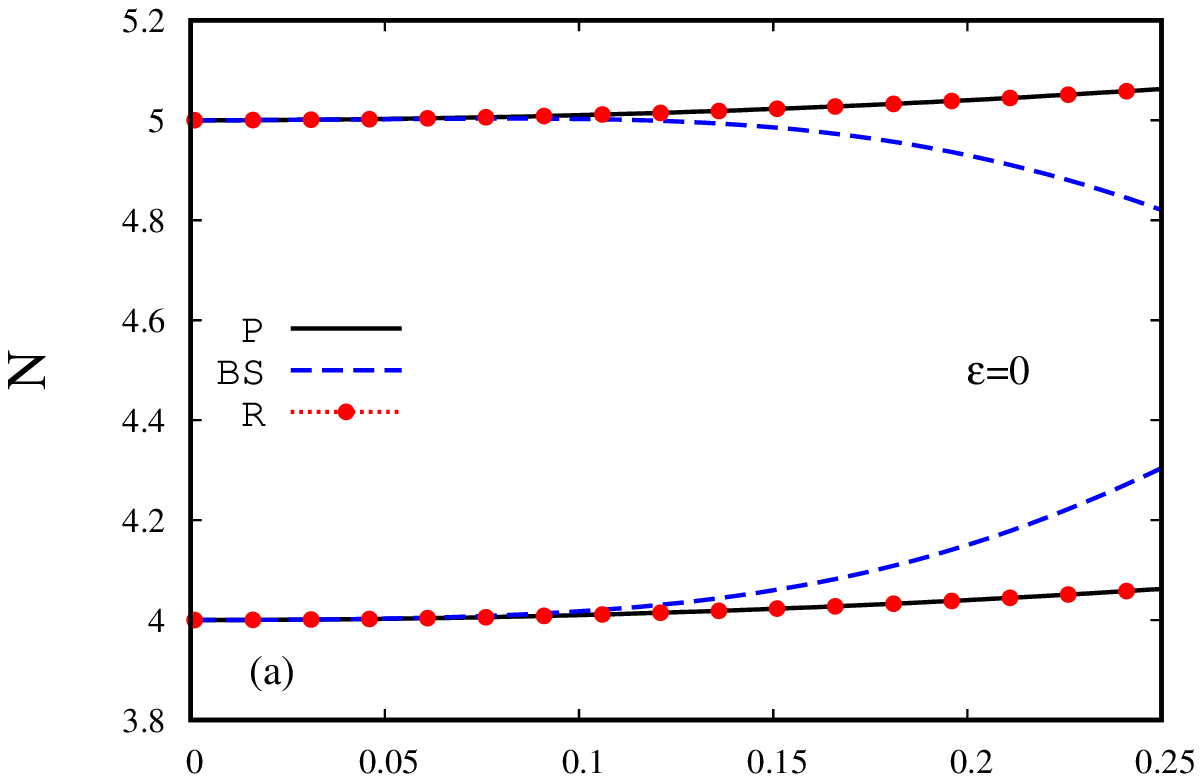}
	\includegraphics[angle=0,width=0.45\textwidth]{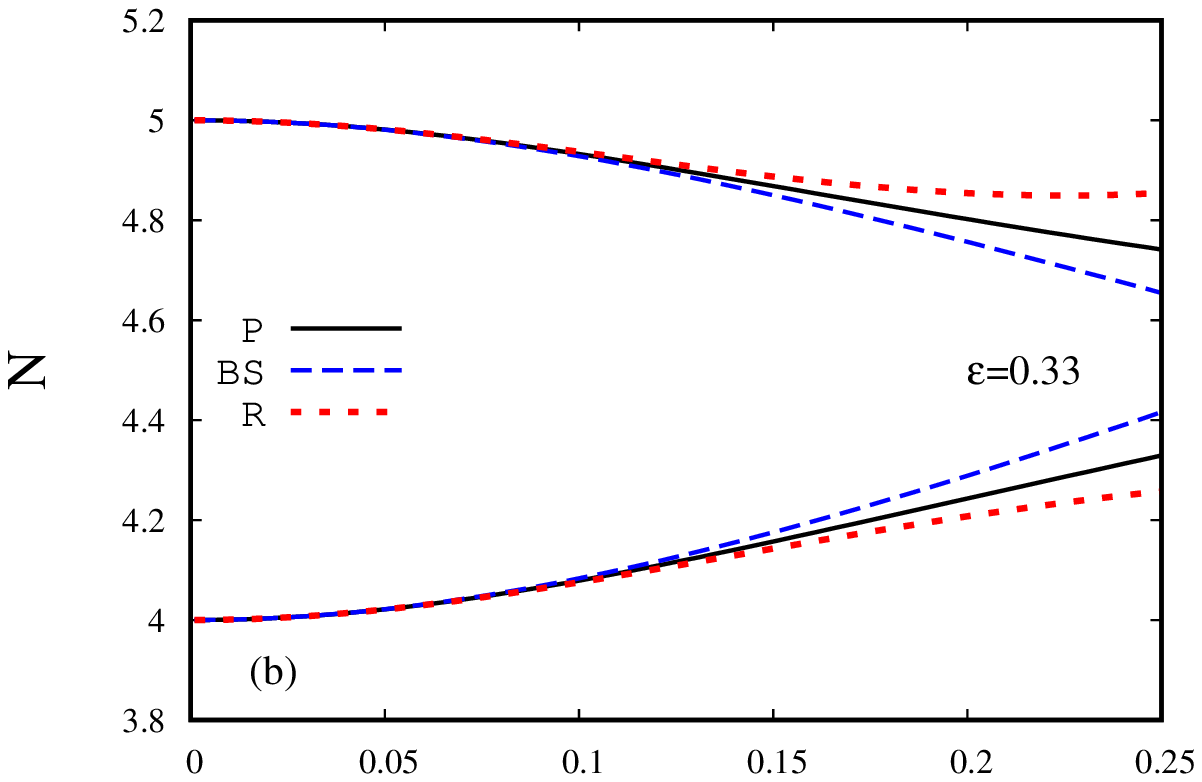}
	\includegraphics[angle=0,width=0.45\textwidth]{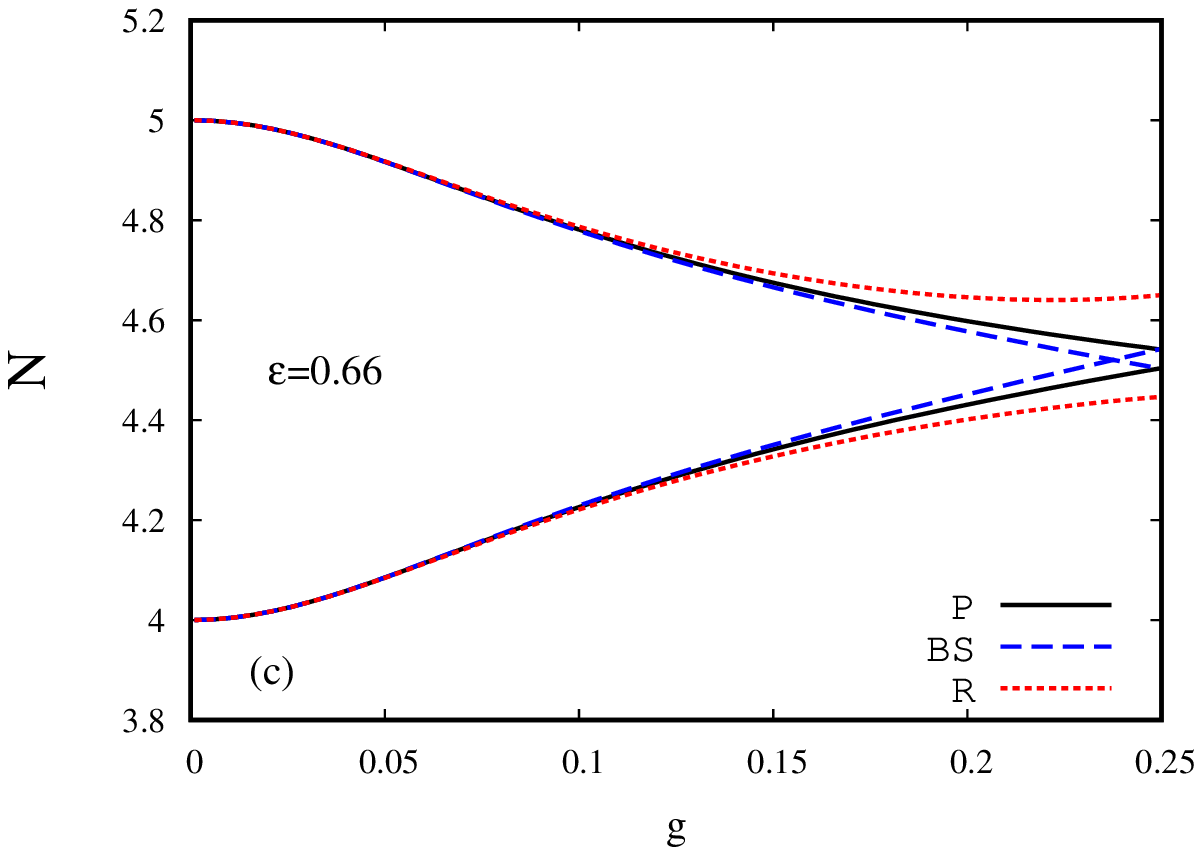}
        \caption{The mean photon number $N$ associated to the eigenstates $|\varphi_{4,\pm}\rangle$ of the Rabi (R), Bloch-Siegert (BS)
        and perturbative (P) Hamiltonians as a function of the matter-photon coupling $g$ at different values of the qubit energy:
         (a) $\ep=0$, (b) $\ep=0.33$ and (c) $\ep=0.66$.}
\label{N-Phot}
  \end{figure}

  The numerical results presented in Fig.\,{\ref{N-Phot}} complement our analysis. We calculated the mean
photon number $N$ corresponding to the eigenstates of the fourth two-dimensional Jaynes-Cummings subspace for the
exact and the effective Hamiltonians. Again, we find that in the IBM limit (see Fig.\,{\ref{N-Phot}}(a)) the
perturbative Hamiltonian reproduces the exact expectation value of the Rabi model (again, we used red circles to
highlight the results of the perturbative model).
Note that for $\ep=0$ one has $S'=0$ too. Thus one is left with only the first transform Eq.\eqref{eq:Sn},
which is exact. Thus not only the Hamitonian, but the 'rotated' form of the $\hat n$ operator is exact.
 In contrast, the results obtained from the Bloch-Siegert model show noticeable differences starting around the qubit-photon
 coupling strenght $g\sim 0.15$.
In particular, the mean photon number corresponding to $|\varphi_{4,+}\rangle$ decreases as $g$ increases.
These differences are even more pronounced for higher two-dimensional subspaces.
As in the case of the spectrum, the photon number expectation values given by the two models become closer as
$\ep$ increases, with the perturbative ones remaining nearer to the exact Rabi values (see Figs.\,{\ref{N-Phot}}(b),(c)).

\subsection{Operator averages. Population inversion and number of excitations}
\label{subsC}
More clearcut differences exist between the two models than suggested by the above 
results, especially at resonance. We show this by performing the same comparison of expectation 
values, this time for the population inversion operator $\si_z$ and for the number of excitations 
$\hat{N}_x=\hat{n}+1/2(1+\sigma_z)$.  

For $H_P$ the change of $\si_z$ under the first unitary transform is obtained as in
the derivation of Eqs.\eqref{eq:Ssi_z},\eqref{eq:no_expansion}.
\begin{align}
    \si_{z,S} =&\, \cosh[2\, \la(a\da-a)]\, \si_z -\sinh[2\, \la(a\da-a)]\,i\si_y
                                       \nonumber \\
      \approx &\,  \si_z + 2\la^2 (a\da-a)^2 \si_z- 2\la(a\da-a) i\si_y \, .                             
\end{align}
The expression remains unchanged by the second transform, which commutes with $\si_z$.
The terms containing $\la$ are left unchanged, as discussed in the previous cases.
Also, as before, the operators $a^2, a^{\dagger\,2}$ and the off-resonant ones 
$a\da\si \da, a \si$ do not contribute to the expectation value. We are left with
\begin{equation}
\si_{z,S,S'}= \,\si_z +2 \la (a\da \si+a\si\da)- 2\la^2 (2a\da a+1) \si_z \, ,
\end{equation}
which by straightforward calculations leads to
\begin{align}
\left<\si_{z}\right>_{P,\pm}& = (u^2_{\pm}-v^2_{\pm})[1-2\la^2(2n+1)] \nonumber \\
                            & +\, 4\la u_{\pm} v_{\pm} \sqrt{n+1} + 4 \la^2 v^2  \, .
\end{align}
Specifically for the two states one obtains

\begin{align}
\left<\si_z\right>_{P,\pm} = &\, \pm\cos\theta_n [1-4\la^2(n+1)] \nonumber \\
                         \pm & 2\la \sin\theta_n\sqrt{n+1} +2\la^2\, .
\label{sig_P}
\end{align}
A negative value for $\cos\theta_n$ hints to a higher population inversion on the lower state.
For the ground state one finds $\left<\si_z \right>_{P,0} = -1 +2\la^2$.

Finally, we address the same problem for $H_{BS}$. The transform of $\si_z$ under 
$\tilde{S}=\la(a\da \si\da-a \si)$ is complicated, but for the second order in $\la$ the 
quantities $[\tilde{S},\si_z]$ and $[\tilde{S},[\tilde{S},\si_z]]$ are sufficient. It is easy to check that
\begin{align} 
[\tilde{S},\si_z] =& \,-2\la (a\da \si\da+a \si) \, ,\nonumber \\
[\tilde{S}, a\da \si\da+a \si] =&\, 2\la(a\da a \si_z - \si\si\da) \, .
\end{align}
This leads to the result
\begin{equation}
\si_{z,\tilde{S}} =\si_z -2\la(a\da\si\da+a\si) -2\la^2 (a\da a \si_z - \si\si\da) \, ,
\end{equation}
which remains unchanged under the second unitary transform. Thus 
\begin{align}
\left<\si_z\right>_{BS,\pm}=&\bra {\varphi_{n,\pm}}
                            [\si_z -2\la^2 (a\da a \si_z-\si\si\da)]
                           \ket {\varphi_{n,\pm}} \nonumber \\
                          =&(u_{\pm}^2-v_{\pm}^2) (1-2\la^2n)+4\la^2 v_{\pm}^2  \, .
\end{align}
Specifically, for the two states one has
\begin{align}
\left<\si_z\right>_{BS,\pm} =&\, \pm\cos\theta_n \,(1-2\la^2n)+2\la^2(1\mp\cos\theta_n)\, \nonumber \\
                            =&\, \pm\cos\theta_n \,[1-2\la^2(n+1)] +2\la^2 \, .
\label{sig_BS}
\end{align}
For the ground state one finds $\left<\si_z \right>_{BS,0} = -1 +2 \la^2$.

The average of the operator $\hat{N}_x$ is now easily done for both effective Hamiltonians by assembling 
Eqs.(\ref{eq:aver_P}), (\ref{eq:aver_BS}),(\ref{sig_P}) and (\ref{sig_BS}). 
The surprising result is that the outcome is the same for both effective models, and reads
\begin{equation}
N_x^\pm =\langle\varphi_{n,\pm}|{\hat N}_x|\varphi_{n,\pm}\rangle =  n + 1 + 2\la^2 \mp 2\la^2(n+1) \cos\theta_n \, .
\label{eq:av_Nx}
\end{equation}

\begin{figure}[t]
        \includegraphics[angle=0,width=0.45\textwidth]{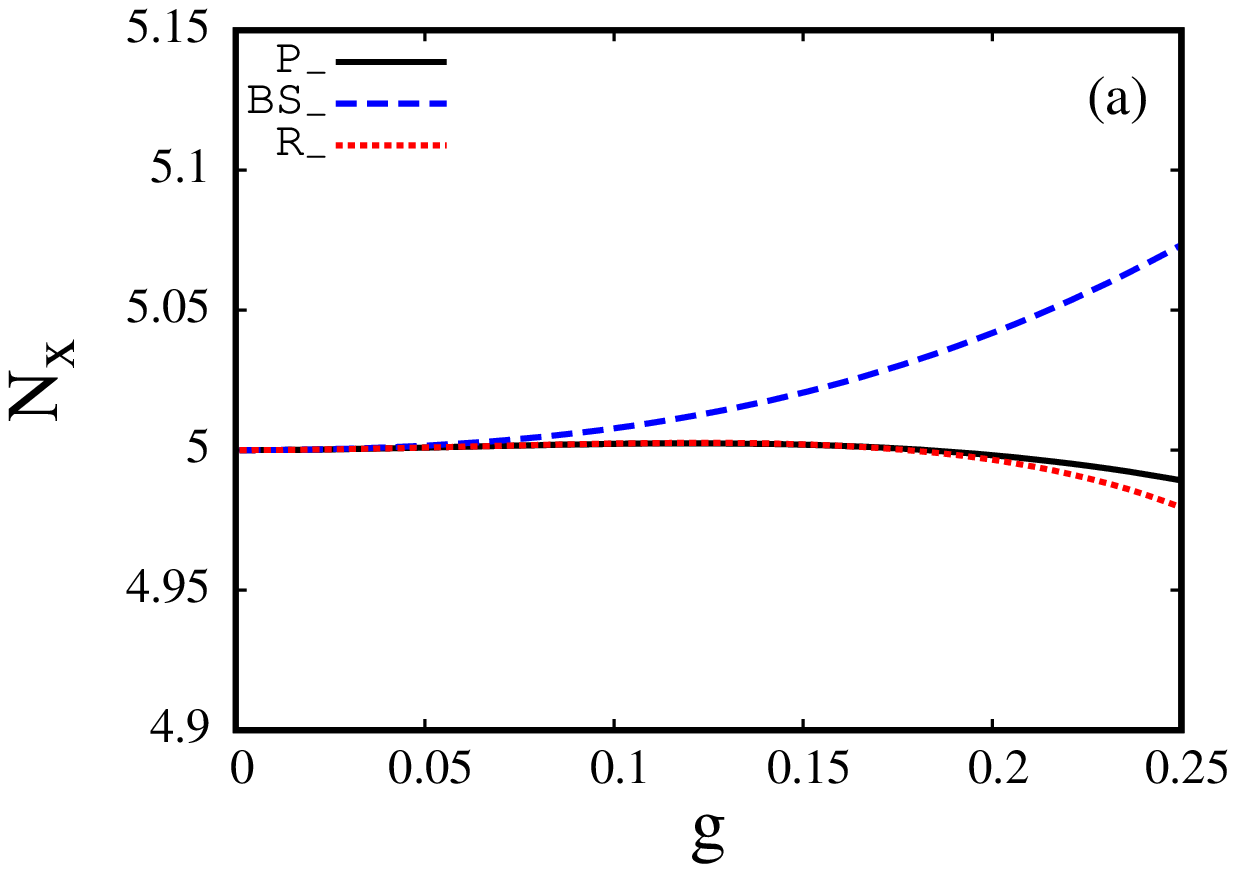}
\hskip -0.55cm
\includegraphics[angle=0,width=0.45\textwidth]{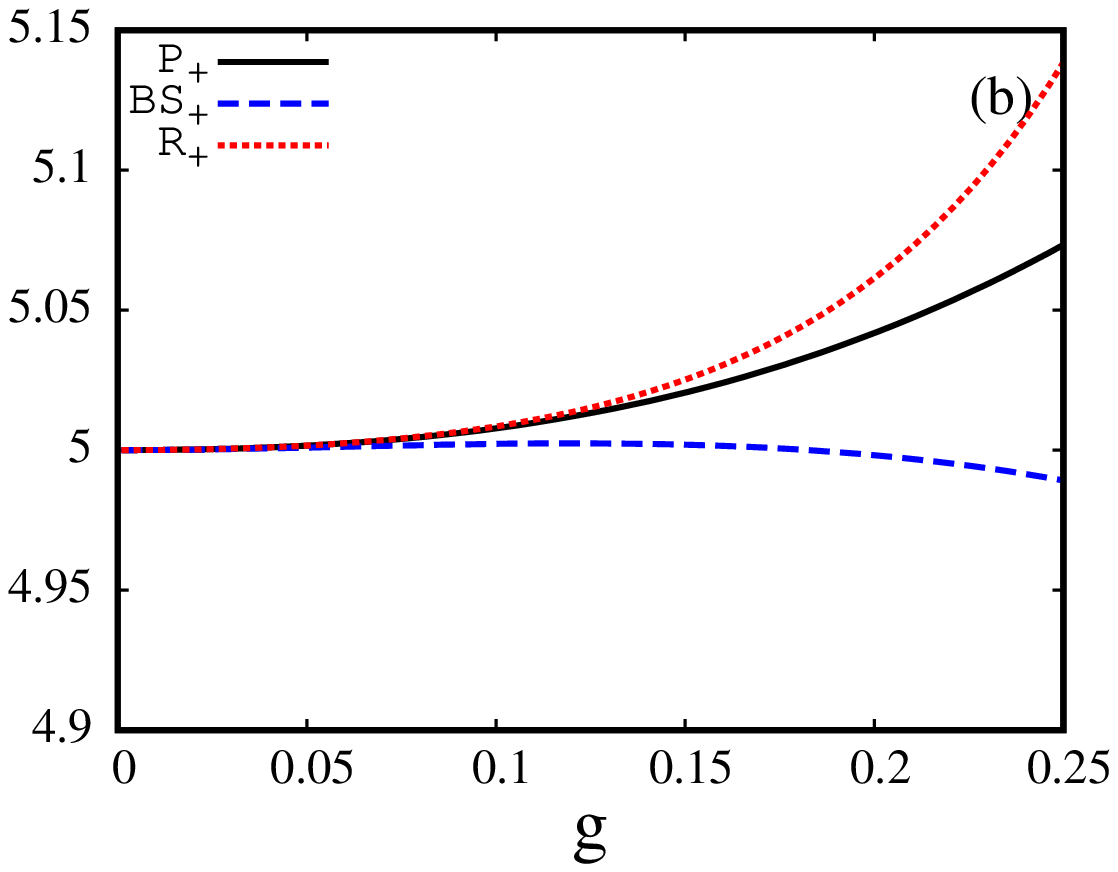}
        \caption{Two branches of the number of excitations $N_x^{\pm}$ associated to the pair of dressed states
        $|\varphi_{4,\pm}\rangle$ as a function of the qubit-photon coupling strenght $g$ in the resonant regime
        $\ep=1$. (a) $N_x^-$, the expectation value of the excitation number for the lower spectral branch;
        (b) $N_x^+$, the same for the upper spectral branch. The obvious similarity between the results of the Rabi
        and the perturbative models confirms the correct assignment of $N_x^\pm$
        to the upper/lower spectral branches.}
\label{NX-branch}
  \end{figure}

This is not a mere coincidence, and relies essentially on the fact that the generators 
of the respective unitary transforms differ by $\bar S:=S-\tilde S = \la (a\da \si- a \, \si \da)$,
which commutes with $\hat N_x$ \cite{Comm-Nx}. 

Again, the interesting situation is provided by the resonant case. There one can see analytically 
that both models predict exactly the same results for a given spectral subspace.
The curves for $N_x^\pm$ are superimposed since, except for the sign, the cosine values coincide.
But because of this sign difference, inherited from the opposite signs of $\Omega_{P}$ and $\Omega_{BS}$,
the assignment of $N_x$ expectation values to the energy branches is switched, $N_{x,BS}^{\pm}=N_{x,P}^{\mp}$. 
More precisely, in the BS model $N_x^-$, corresponding to the lower branch, takes the higher value, 
while $N_x^+$, stemming from the upper branch, is smaller. In the perturbative case, it is the other way around. 
This is an important difference, expressing the distinct structure of the eigenstates, and which cannot 
gradually disappear as $\la$ gets smaller. To decide which attribution is the correct one we rely on comparison 
with the result of the full Rabi Hamiltonian.

The outcome is shown in Fig.\,\ref{NX-branch}, in which the values of the two models are plotted, along with 
the numerical calculation using the dressed states provided by diagonalization of the Rabi Hamiltonian.
The left panel corresponds to the eigenstates $\ket{\varphi_{4,-}}$ and the right one to $\ket{\varphi_{4,+}}$.
One can check by inspection of  Figs.\,\ref{NX-branch} (a) and (b) that $N_{x,P}^{\pm}=N_{x,BS}^{\mp}$,
and that comparison with the Rabi solutions decides in favor of the perturbative results.

\begin{figure}[t]
        \includegraphics[angle=0,width=0.45\textwidth]{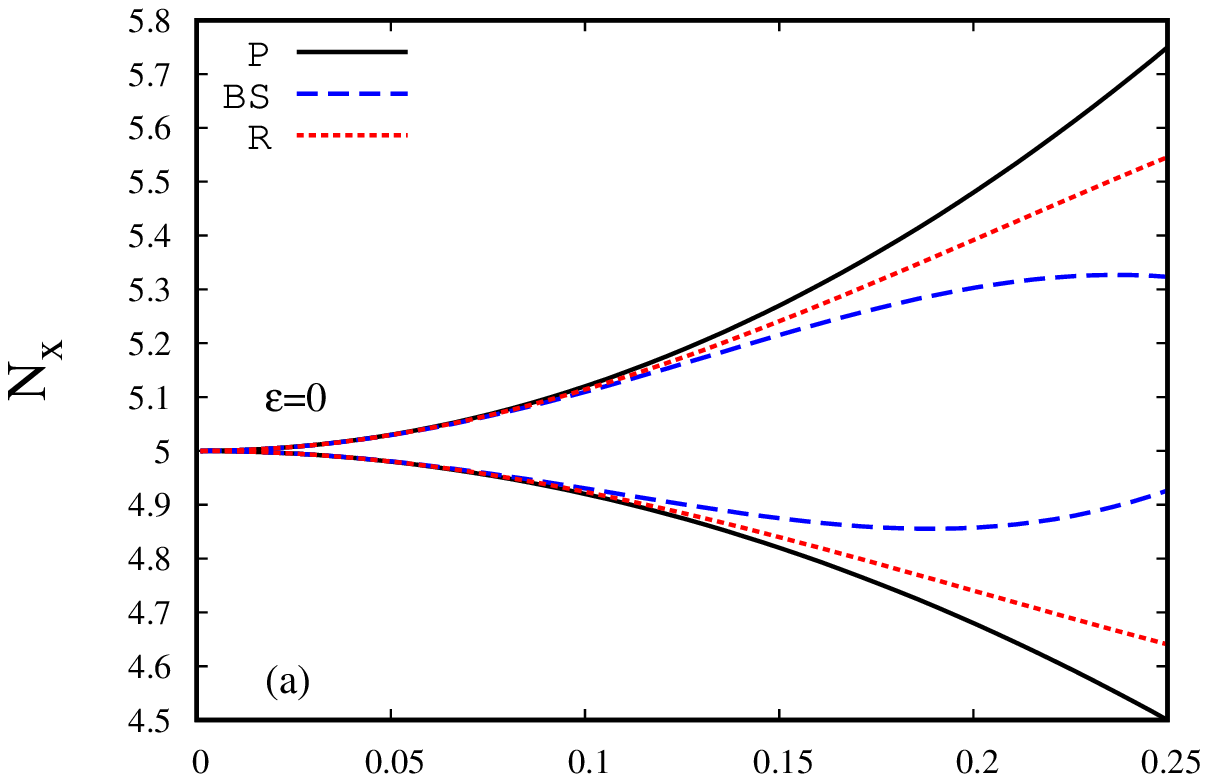}
	\includegraphics[angle=0,width=0.45\textwidth]{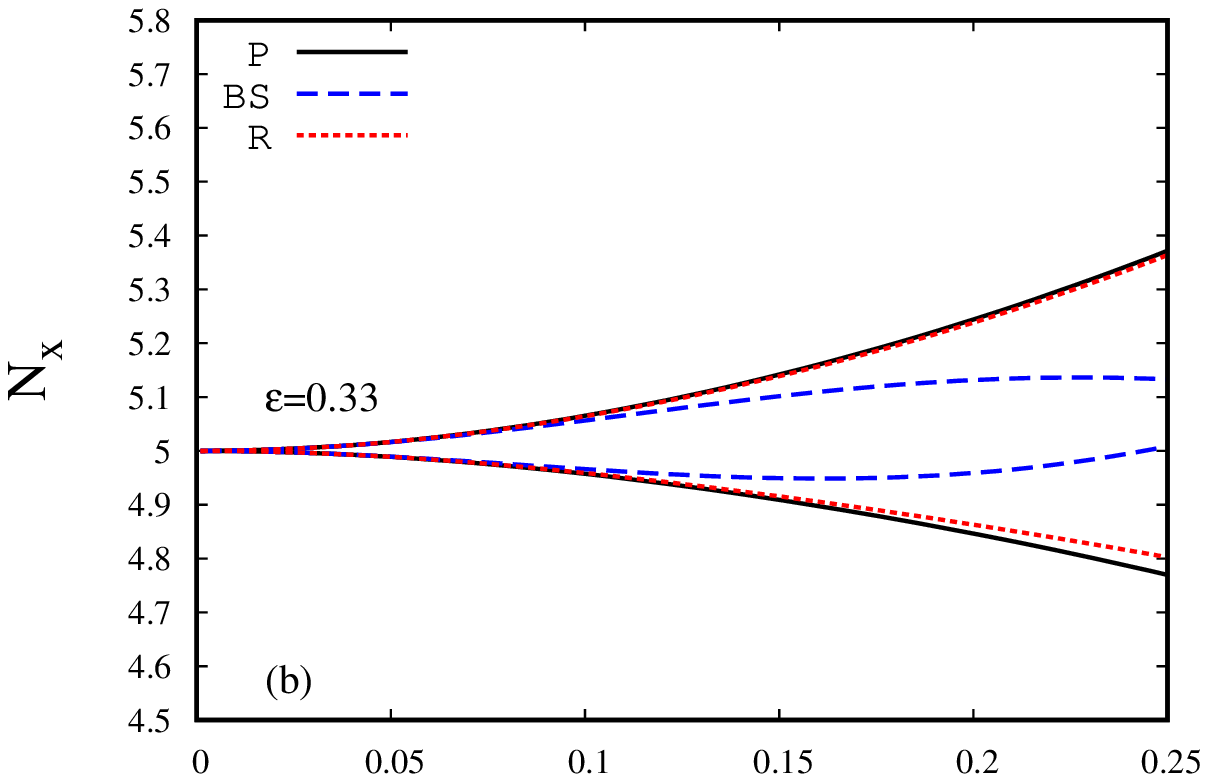}
	\includegraphics[angle=0,width=0.45\textwidth]{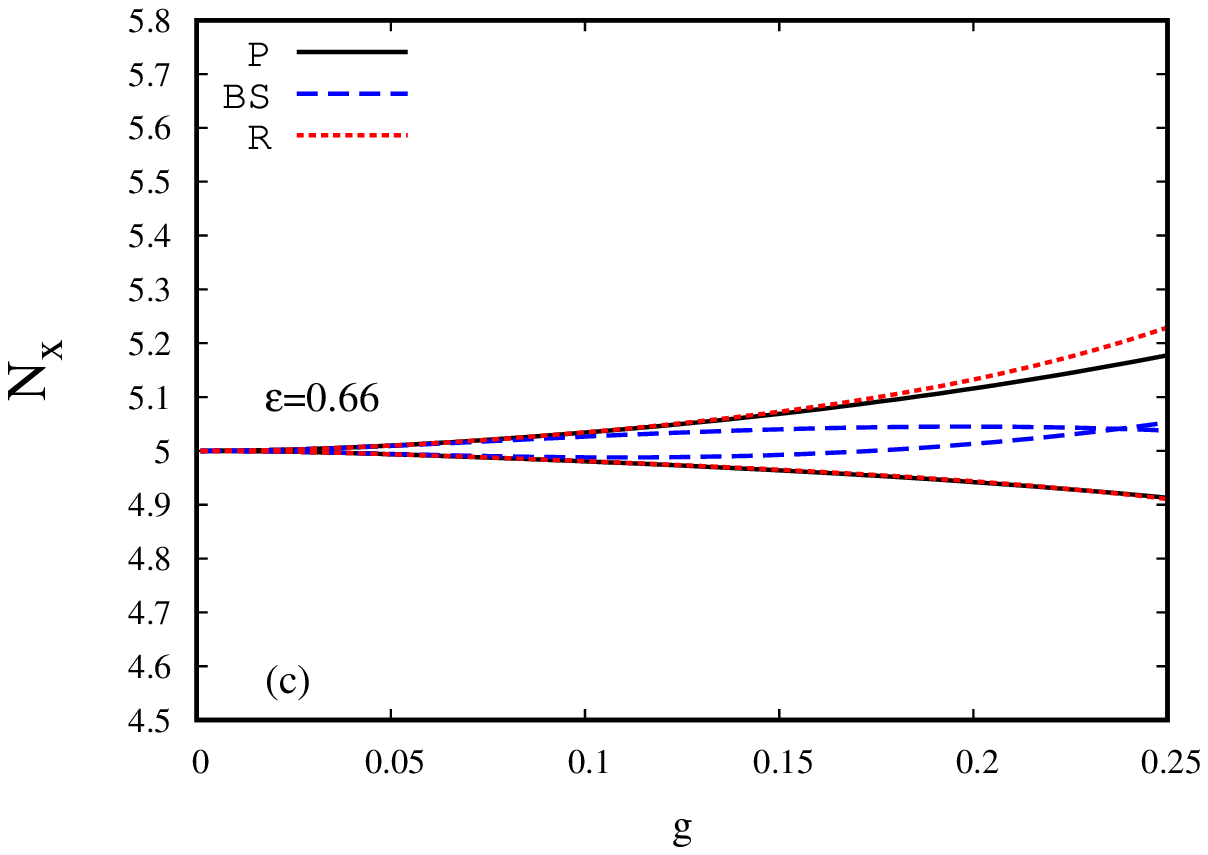}
\caption{The number of excitations $N_x$ associated to the pair of dressed states in
        the fourth spectral subspace as a function of the field-matter coupling strength $g$,
        for the perturbative (P), Bloch-Siegert (BS) and Rabi (R) Hamiltonians.
        (a) The IBM-limit $\ep=0$, (b) $\ep=0.33$, (c) $\ep=0.66$.}
\label{NX1}
  \end{figure}

Finally we compare in Fig.\,\ref{NX1} the number of excitations for the same values of the qubit energy considered in
previous discussions. In the IBM-limit $N_{x,P}^{\pm}$ follow qualitatively the exact curves but also deviate from
them as $g$ increases. (Let us stress that for $\ep=0$ the exact result is analytically available in terms of Laguerre 
polynomials \cite{Mahan}, but here we stick to Eq.\eqref{eq:av_Nx}, in which the operators are rotated only to second order).
On the other hand, the excitation numbers of the BS model are quite far from the
$N_{x,R}^{\pm}$ for all values of $\ep$. In particular, at $\ep=0.66$ the branches of $N_{x,BS}^{\pm}$ exhibit an incorrect 
crossing point around $g=0.235$. Moreover, the BS attribution of the excitation number to a given spectral branch is still 
flipped and the inequality $N_{x,BS}^{+}<N_{x,BS}^{-}$ still holds. On the contrary, the branches $N_{x,P}^{\pm}$ follow 
closely the exact ones in the sub-resonant regime too.

\subsection{The generalized RWA effective Hamiltonian}

A successful effective Hamiltonian suitable in the deep strong coupling (DSC) regime defined by $g/\om \geqslant 1\,$ \cite{RMP} was 
proposed by Irish \cite{Irish}. This so called generalized RWA (GRWA) Hamiltonian follows from $H_R$ (see Eq.\eqref{eq:hqrm}) by the 
IBM transform. The free qubit part of the Hamiltonian then couples all pairs of shifted oscillator states.  

To be specific, the model uses a single canonical transform, generated by 
$S=\la(a\da-a)\si_x$, this time with $\la = g/\om$. As a result one obtains, as 
for Eq.\eqref{eq:no_expansion}
\begin{align}
& H_{R,S} =  \om a\da a - g^2/\om  \nonumber \\
& +\, \frac{1}{2} \ep \left \{\cosh[2\, \la(a\da-a)]\, \si_z 
                    -\sinh[2\, \la(a\da-a)]\,i\si_y \right \} \,. 
  \label{eq:irish}
\end{align}
It is lengthy but rather straighforward to expand the hyperbolic functions as
the even and odd part of the exponential (see Xie {\it et al.} \cite{Xie} but 
beware of typos)
\begin{align}
e^{2\,\la (a\da -a)} & = f_0(\hat{n},\la) + \nonumber \\  
            & \sum_{q>0} \left[a^{\dagger\, q}f_q(\hat{n},\la)
            +(-1)^q f_q (\hat{n},\la) a^q\right] \, ,
\end{align}
where the functions of $\hat{n}$ are defined, as usual, by their action 
on the elements of the Fock basis $f_q (\hat{n},\la)\ket{n} =f_q (n,\la)\ket{n}$. 
They are related to the generalized Laguerre polynomials $L_n^{(q)}(x)$ by
\begin{equation}
f_q (n, \la) = e^{-2\la^2}(2\la)^q\,\frac{n!}{(n+q)!}\, L^{(q)}_n (4\la^2) \, . 
\label{eq:laguerre}
\end{equation}
This translates into the following expression for $H_{R,S}$ 
\begin{align}
H_{R,S} = \,& \om \, a\da a - g^2/\om + \frac{\ep}{2} f_0 (\hat{n},\la) \si_z \nonumber \\ 
          & + \frac{\ep}{2} \sum_{q>0, \rm{even}} \,
                  \left[a^{\dagger\, q}f_q(\hat{n},\la)
                     + f_q(\hat{n},\la) a^q\right] \}\, \si_z \nonumber \\
          & + \frac{\ep}{2} \sum_{q>0, \rm{odd}}\,\left[a^{\dagger\, q}f_q (\hat{n},\la)
                     - f_q(\hat{n},\la) a^q\right] (\si - \si \da) \, .
\end{align}
No approximation is involved so far. The $f_0$ term is diagonal, while terms with 
higher $q$-values are further away from the diagonal in the Fock basis. 

The idea of the GRWA is to keep, besides the diagonal terms, only the off-diagonal 
ones with $q=1$ and which, moreover, are also 
retained in the RWA, as not counter-rotating. More precisely, one has  
\begin{align}
H_{{\rm GRWA}} = &\,\om \, a\da a - g^2/\om 
                   + \, \frac{\ep}{2}\, f_0(\hat{n},\la)\, \si_z \nonumber \\
                   + & \, \frac{\ep}{2} \, [a \da \si f_1(\hat{n},\la)
                                      + f_1(\hat{n}, \la)\, a \si \da] \, , 
                                      \label{eq:hgrwa}
 \end{align}
which reduces the Hamiltonian to the familiar JC structure, renormalized by 
$\hat{n}$-dependent factors. Thus the problem becomes $2 \times 2$ block-diagonal and 
therefore solvable. 

Obviously, this approach is not a systematic expansion to $\la^2$. Indeed, $f_q \sim \la^q$
and therefore terms with $f_2$ should have been kept as second order contribution. Instead 
they were discarded as off-resonant, as prescribed by the RWA argument. 

\begin{figure}[t]
         \includegraphics[angle=0,width=0.45\textwidth]{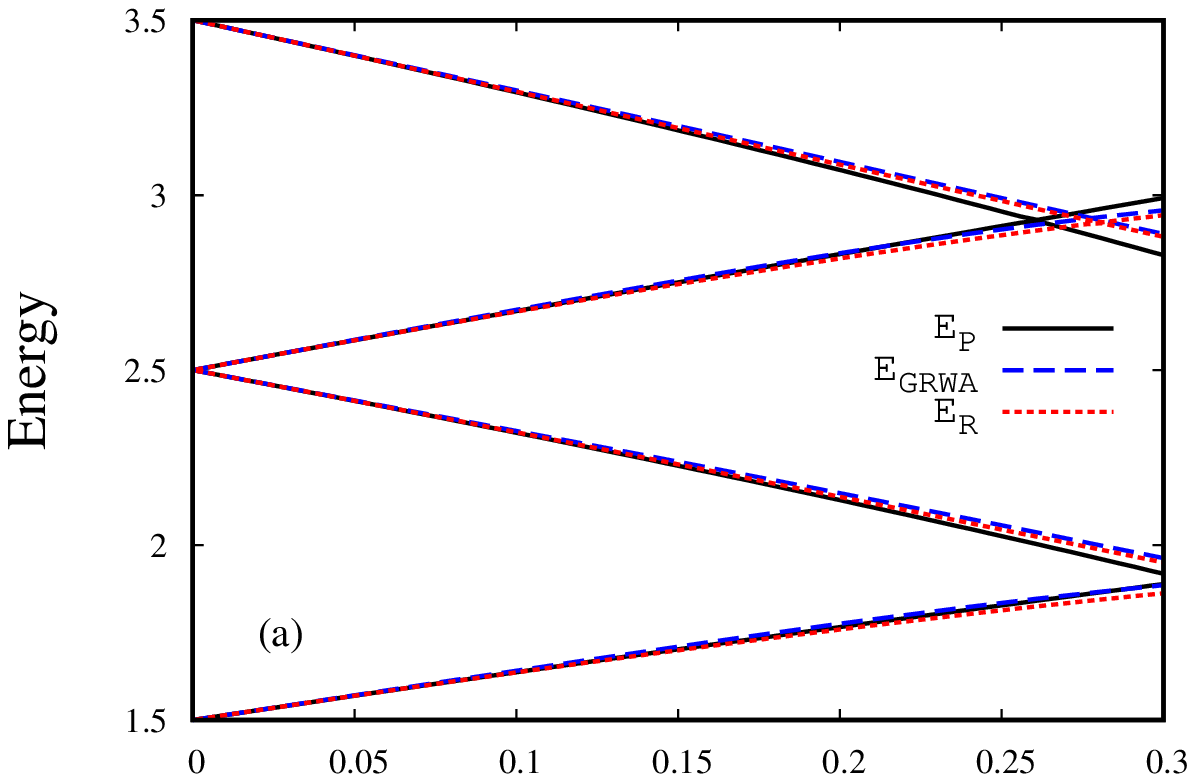}
         \includegraphics[angle=0,width=0.45\textwidth]{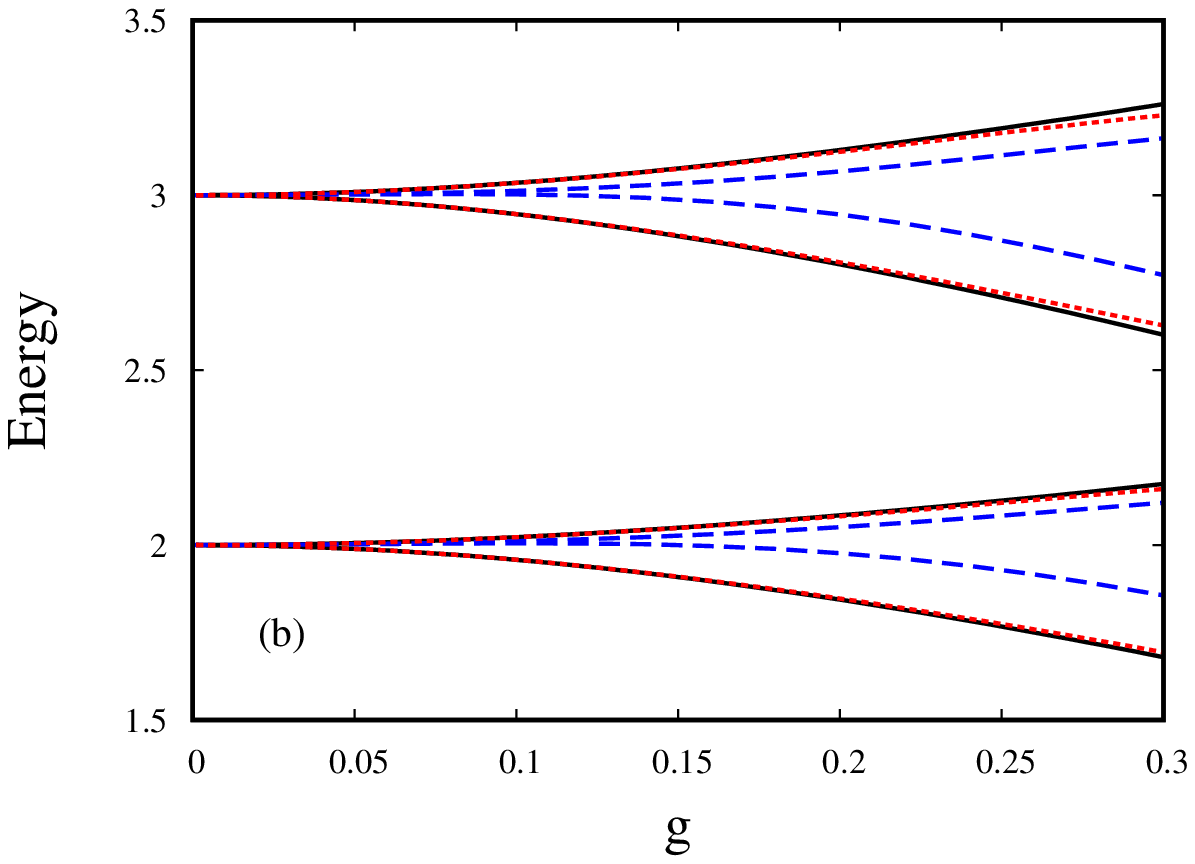}
         \caption{Several spectral branches of the Rabi (R), GRWA and perturbative (P) Hamiltonians
 as a function of the matter-photon coupling $g$ in the (a) resonant regime $\ep=1$ and (b) supra-resonant regime 
 $\ep=2$. $g$ is restricted to the perturbative USC regime.}
 \label{Comp-Irish}
   \end{figure}

The success of the model in the DSC regime relies on the fact that {\it all} terms arising
from the transformed qubit energy become small at large $g$, due to the exponential prefactor.
Suprisingly, its predictions work quite well at lower couplings too. This means that at least
in certain circumstances the off-resonant terms are negligible indeed. On the contrary, in situations like
$\ep \approx 2\om$ this is not the case any more. Terms like $a^{\dagger\,2} \si$ now describe resonant transitions
in the system and thus are no longer negligible. In contrast, these processes are included in the present perturbative
approach by its second unitary transform $S'$.

The situation is illustrated in Fig.\,\ref{Comp-Irish} where we compare a part of the low-energy spectra of the Rabi,
GRWA and perturbative Hamiltonians
in the resonant ($\ep=\om$)  and supra-resonant ($\ep>\om$)  regimes. The eigenvalues of the GRWA Hamiltonian \cite{Irish}
are given analytically by Eq.\eqref{eq:hgrwa}. Clearly, the eigenvalues of both GRWA and perturbative
Hamiltonians reasonably follow the Rabi spectrum in the resonant case along the whole range of the pUSC regime
(see Fig.\,\ref{Comp-Irish}(a)). However, Fig.\,\ref{Comp-Irish}(b) reveals that in the supra-resonant regime
the GRWA spectrum is no longer accurate in the pUSC regime while the errors of the perturbative model are quite small.

\section{Conclusions}

The effective Hamiltonian given by Eq.\eqref{eq:heff2} was derived from the quantum Rabi model systematically 
to second order in the small parameter $\la=g/(\ep+\om)$. No other approximations (like discarding
terms as off-resonant) are involved. In this sense the proposed model is purely perturbative.
The expansion parameter $\la$ is the same as in the Bloch-Siegert theory. In contrast to the latter, 
the present formalism recovers exactly the IBM Hamiltonian in the limit $\ep \to 0$. 
The proposed model allows analytical calculations of various operator averages 
which are used for a detailed comparison with the predictions of the BS approach. 

Even in the cases when the spectra of two models are close, we find that the main difference from the BS results 
is seen in the structure of the dressed eigenstates. The source of this dissimilarity is traced back to the different 
sign of the frequency shift induced by the qubit-field coupling. This is furher reflected in  the expectation values 
of relevant observables, like the excitation number, and their attribution to the appropriate eigenstates. 

The energy spectrum and the dressed state structure associated to 
the full quantum Rabi model are well approximated by the IBM-compatible effective Hamiltonian in the resonant 
and off-resonant perturbative USC regime. 
Our analysis shows that recovering the independent-boson model in the limit 
$\ep\to 0$ is a valuable test for effective Hamiltonian approaches to the quantum Rabi model.

\begin{acknowledgments}
V.M. acknowledges the financial support from the Romanian Core Program PN19-03 (contract no.\ 21 N/08.02.2019).
\end{acknowledgments}


\begin{thebibliography}{20}
\bibitem{Comm-Nx} 
Other technicalities play a role too: the commutator $[\tilde S, \bar S]$ generates only off-diagonal terms,
which do not influence the $\hat N_x$ expectation value.
\end{thebibliography}
\end{document}